\pdfoutput=1

\documentclass[11pt]{article}

\usepackage{acl}

\usepackage{times}
\usepackage{latexsym}

\usepackage[T1]{fontenc}

\usepackage[utf8]{inputenc}

\usepackage{microtype}

\usepackage{inconsolata}

\usepackage{subfigure}
\usepackage{graphicx}
\usepackage{booktabs}
\usepackage[most]{tcolorbox}
\usepackage{listings}
\newcommand{\code}[1]{\lstinline|#1|}

\title{

A Distributed Collaborative Retrieval Framework Excelling in All Queries and Corpora based on Zero-shot Rank-Oriented Automatic Evaluation

}

\author{
\parbox{0.8\linewidth}{
\centering
  Tian-Yi Che$^1$\hspace{1em}
  Xian-Ling Mao$^1$\hspace{1em}
  Chun Xu$^{1,2}$\hspace{1em}
  Cheng-Xin Xin$^1$\hspace{10em}
  Heng-Da Xu$^1$\hspace{1em}
  Jin-Yu Liu$^1$\hspace{1em}
  Heyan Huang$^1$}\vspace{0.12cm}\\
  $^1$School of Computer Science \& Technology, Beijing Institute of Technology \\
  $^2$School of Mathematics and Computer Science, Yan'an University \\
\texttt{\{ccty, maoxl, xin.chengxin, xuhengda, hhy63\}@bit.edu.cn} \\
\texttt{xuchun@yau.edu.cn}, \hspace{1em} \texttt{liujinyu1229@gmail.com}
}
\vspace{-0.5cm}

\begin{document}
\maketitle
\begin{abstract}
Numerous retrieval models, including sparse, dense and llm-based methods, have demonstrated remarkable performance in predicting the relevance between queries and corpora. However, the preliminary effectiveness analysis experiments indicate that these models fail to achieve satisfactory performance on the majority of queries and corpora, revealing their effectiveness restricted to specific scenarios. Thus, to tackle this problem, we propose a novel \textbf{D}istributed \textbf{C}ollaborative \textbf{R}etrieval \textbf{F}ramework (\textbf{DCRF}), outperforming each single model across all queries and corpora. Specifically, the framework integrates various retrieval models into a unified system and dynamically selects the optimal results for each user's query. It can easily aggregate any retrieval model and expand to any application scenarios, illustrating its flexibility and scalability.
Moreover, to reduce maintenance and training costs, we design four effective prompting strategies with large language models (LLMs) to evaluate the quality of ranks without reliance of labeled data. 
Extensive experiments demonstrate that proposed framework, combined with 8 efficient retrieval models, can achieve performance comparable to effective listwise methods like RankGPT and ListT5, while offering superior efficiency. Besides, DCRF surpasses all selected retrieval models on the most datasets, indicating the effectiveness of our prompting strategies on rank-oriented automatic evaluation.

\end{abstract}

\section{Introduction}

\begin{figure}[t]
    \centering
    \includegraphics[width=\linewidth]{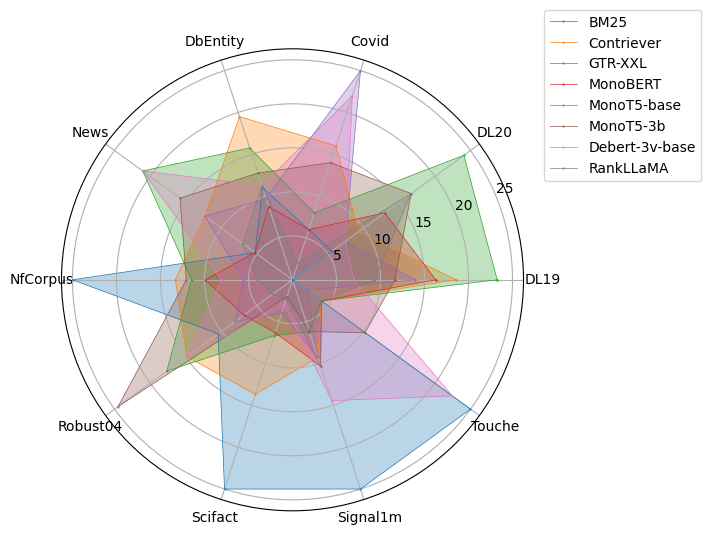}
    \caption{The frequency of retrieval models on the benchmark TREC and BEIR. The frequency means the proportion of queries on which a specific model generates the best results among all selected models. The \% of values is omitted and the maximum displayed value is 25\% for legibility. A region denotes the capacity of a retrieval model across all the queires and corpora and the most ideal shape is the ten-sided polygon.}
    \label{fig1}
\end{figure}

\begin{figure*}[t]
    \centering
    \subfigure[Traditonal retrieval-rerank framework]{
        \includegraphics[width=0.65\linewidth]{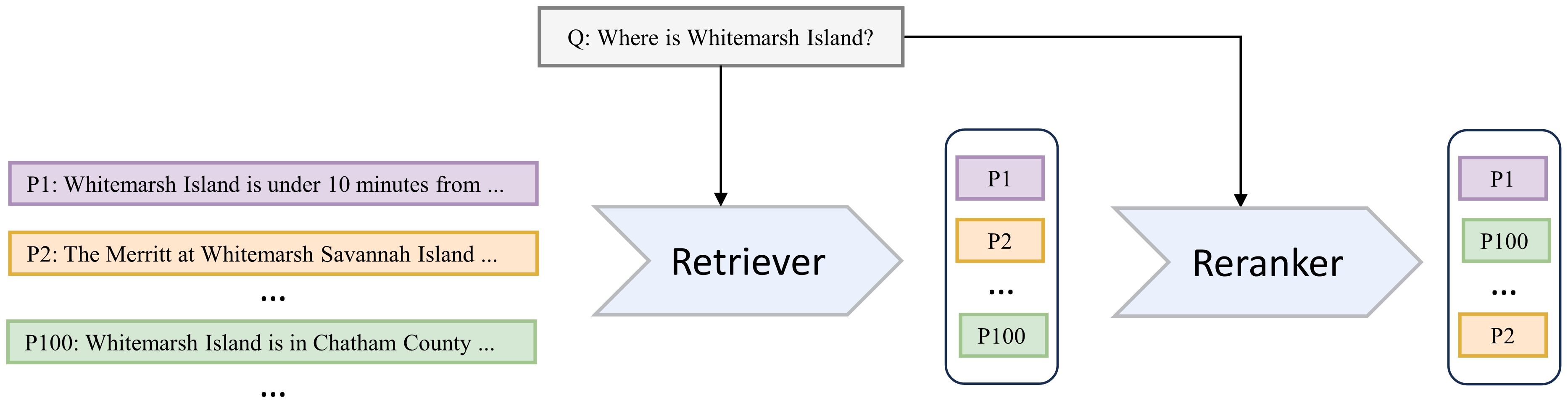}
    }
    \subfigure[Proposed retrieval-rerank-evaluation framework]{
        \includegraphics[width=0.95\linewidth]{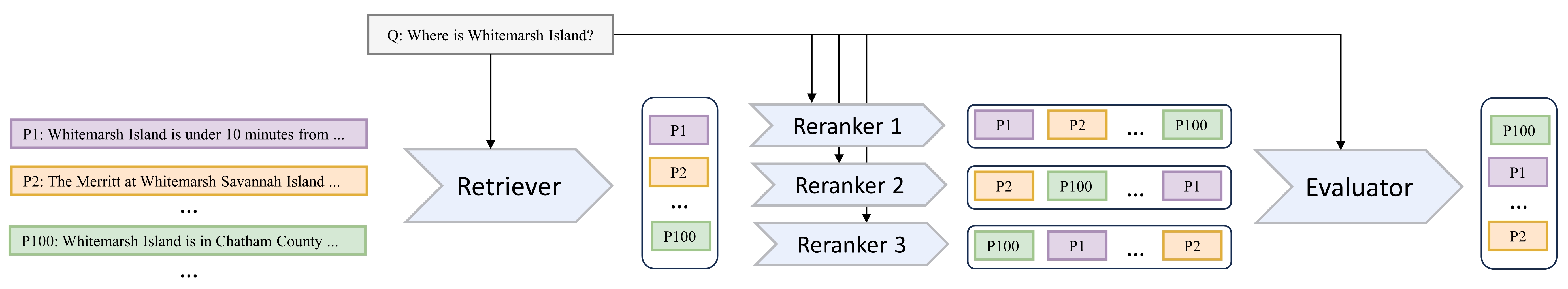}
    }
    \caption{The comparison of the traditional IR framework and DCRF. (a) A retriever is used to recall top-100 relevant passages and a reranker reorders them based on their relevancy to the query. (b) The retriever is the same and then multiple rerankers reorder the recalled passages respectively as candidates. Finally, the evaluator picks up the best rank from the candidates based on the query.}
    \label{fig2}
\end{figure*}

Information Retrieval (IR) is a crucial and valuable research task, aiming to help users quickly capture their desired information. Over the past years, a lot of effective methods have been proposed to improve the performance of retrieval from sparse models to dense models and then to LLMs \cite{Zhao2022DenseTR, Zhu2023LargeLM, Wang2024UtilizingBF}. Specifically, at first, researchers focus on studying term-based methods, such as TF-IDF and BM25 \cite{Robertson2009ThePR}, which are the foundation of search engines. Next, as the improvement of pretrained language models, a number of excellent works based on representation learning have emerged. The typical works are DPR \cite{Karpukhin2020DensePR}, ColBERT \cite{Khattab2020ColBERTEA} and Contriever \cite{Izacard2021UnsupervisedDI}. Nowadays, LLMs have demonstrated strong generalization on downstream NLP tasks and there are many investigations to study how to employ LLMs for ranking tasks, such as RankGPT \cite{Sun2023IsCG}.

However, the experimental results in their papers indicate that even the start-of-the-art model, RankGPT, can just have the optimal performance on half of the benchmarks, i.e., 5 out of 10 datasets. Inspired by that, we make the effectiveness analysis experiments to evaluate the models' performance across all queries and corpora in the benchmark TREC \cite{Craswell2021OverviewOT} and BEIR \cite{Thakur2021BEIRAH}. The overall results and the definition of frequency are illustrated in Figure \ref{fig1}. It can be observed that no model is able to achieve good frequency on more than three datasets, falling far short of the ideal state represented by the ten-sided polygon. Therefore, it is highly valuable to develop a framework that can dynamically select the best model based on the user's query, in order to adapt to all the queries and corpora.

In this paper, we propose a novel Distributed Collaborative Retrieval Framework (DCRF) that can integrate various retrieval models into a unified system and dynamically select the optimal results for each user’s query, as shown in Figure \ref{fig2}. The largest technical challenge is the design of an effective evaluator, whose quality will significantly affect the performance of the framework. Considering the maintenance and training costs, we design four effective prompting strategies with large language models (LLMs) to evaluate the quality of ranks without reliance of labeled data. This framework is highly flexible and extensible, enabling the integration of any new model to accommodate emerging scenarios. In theory, DCRF can achieve the strongest performance among all retrieval models and we will remain dedicated to designing robust evaluation methods to arouse its capabilities.

Following the RankGPT, we choose 10 datasets in the benchmark TREC and BEIR. To validate the adaptability and generalization of the prompting evaluation strategies, we leverage the prompting strategies on 6 open-source LLMs with different structure, training data and parameter quantity.
First of all, the effectiveness and potential analysis experiments simulate the optimal effect of DCRF with annotated references, demonstrating that this framework has excellent upper bound of performance. Second, the comparison with selected retrieval models and other state-of-the-art methods indicates the effectiveness of the framework and prompting strategies. Third, the ablation experiments indicate that LLMs with more parameters have better performance and the passage-pointwise strategy is the most suitable method for them. Finally, we analyze two important problems and get the following conclusions: (1) the parallel DCRF has lower inference time cost compared to the listwise methods, such as RankGPT and ListT5. (2) the positional bias of the proposed prompting evaluation strategies is minimal.

\section{Related Works}

\subsection{Retrieval Models}
The retrieval task aims to recall and rank the passages based on the relevence to the query, such as BM25 and Contriever \cite{Izacard2021UnsupervisedDI}.
In the retrieval scenario, rather than dual-encoder models \cite{Karpukhin2020DensePR} that separately encode the passages and queries, models that see query and passage information jointly in inference time \cite{Nogueira2019PassageRW, Nogueira2020DocumentRW} are shown to be effective for zero-shot retrieval \cite{Rosa2022InDO}.
Among those, formulating retrieval as sequence generation, such as conducting listwise sorting \cite{Sun2023IsCG, Pradeep2023RankZephyrEA, Yoon2024ListT5LR} or generating rationales \cite{Ferraretto2023ExaRankerEN}, has shown an advantage in application to zero-shot retrieval by leveraging the language model’s auto-regressive generation capabilities. 
However, all of these methods utilize only one model to rank the passages across all scenarios, leading to the limitation of models' applicability and generalization.

\subsection{LLMs for Evaluation}

Automatic evaluation is a task aiming to remove the reliance on manually written reference texts \cite{scialom2019answers, vasilyev2020fill, rei2021references}. Following the line of work, recent studies on LLM evaluators have shown that LLMs can serve as high-quality evaluators for various NLP tasks \cite{fu2023gptscore}, including Summarization \cite{Shen2023LargeLM}, Machine Translation \cite{Kocmi2023LargeLM} and Factual Consistency Evaluation \cite{Luo2023ChatGPTAA}. However, they have reported that LLM evaluators also have numerous biases, such as positional bias \cite{Wang2023LargeLM}, verbosity bias \cite{Zheng2023JudgingLW} and self enhancement bias \cite{Panickssery2024LLMER}. Following these studies, we focus on the rank-oriented LLM  evaluation methods whose target isn't a sequence but an ordered list of sequences.

\section{Proposed Framework}


As illustrated in Figure \ref{fig2}, different from the traditional two-stage IR framework, the proposed DCRF involves three stages, including retrieval, rerank and evaluation.
The first two stages are the same as the previous framework, aiming to recall top-k relevant passages and re-rank them to an ordered result. The third stage is dedicated to evaluate the quality of these ranked results in relation to the user's query and select the optimal result for return.
Following previous work \cite{Sun2023IsCG}, we utilize the BM25 algorithm as the first-stage retrieval and recall 100 passages for subsequent ranking. Next, we will introduce the design of the rerank and evaluation in detail.

\subsection{Criterion of Reranker Selection}
\label{sec:base_retrieval models}

The comprehensiveness and diversity of rerankers significantly affect the effectiveness and efficiency of the framework. Thus, inspired by previous works \cite{Zhao2022DenseTR, Zhu2023LargeLM}, we classify existing retrieval models into three types, including sparse retrieval, dense retrieval and llm-based retrieval. They exhibit differences in terms of architecture, modeling approaches and training data, which are tailored to suit various application scenarios. Besides, all the rerankers are conducted in parallel, meaning that the inference time is determined solely by the slowest reranker, thereby ensuring the efficiency of the distributed framework.

\subsubsection{Sparse Retrieval}
Sparse models in information retrieval are designed to rank documents based on the occurrence of query terms within the document and their distribution across the entire document collection. We select the BM25 as one reranker, because it extends the classical term frequency-inverse document frequency (TF-IDF) model by incorporating term frequency saturation and document length normalization, which improves its robustness and effectiveness in information retrieval systems. Even compared to dense and llm-based retrieval, BM25 still has outstanding performance in long document scenarios, so it's crucial to select it as one reranker to improve the effectiveness of DCRF.

\subsubsection{Dense Retrieval}

Dense retrieval methods represent a shift from traditional sparse retrieval models by leveraging dense vector representations of both queries and documents. They can be usually categorized into three main types: (1) \textbf{Bi-encoders} that separately encode the query and passage into a sentence embedding and then calculate the similarity of two embeddings. (2) \textbf{Cross-encoders} that input the concatenation of the query and passage into one encoder to predict their relevancy between 0 and 1. (3) \textbf{LM-based models} that aim to predict the probability that next token is true or false based on the query and passage.

In order to ensure the diversity of the framework, we choose the typical models in each type of dense retrieval methods. First, in the Bi-encoders, the BERT-based Contriever \cite{Izacard2021UnsupervisedDI} and T5-based GTR \cite{Ni2021LargeDE} are selected because of the strong generalization from the contrastive pre-training on large-scale corpus. They both utilize the mean pooling to capture the vector representation of the whole passage, which means the preference of the short passages. Besides, in the Cross-encoders, we select the MonoBERT \cite{Nogueira2019PassageRW}, which has the optimal performance in this modeling methods. It utilizes the [CLS] vector of BERT model as input to a single layer neural network to obtain the probability of the passage being relevant. Finally, in the LM-based models, MonoT5 \cite{Nogueira2020DocumentRW} is the best-known retrieval model, which is usually chosen as the baseline of retrieval tasks. Meanwhile, considering the effect of parameter quantity on models' preference, we select both the MonoT5-base and MonoT5-3b as the rerankers.

\subsubsection{LLMs for retrieval}

Unlike traditional retrieval models that rely on keyword matching or ranking, LLM-based retrieval focuses on generating natural language responses or summaries based on the given query and the candidate passages. Specifically, the llm-based retrieval can be divided into pairwise, pointwise and listwise methods. The pairwise methods compare the relevance to the query between pairs of passages, lacking the efficiency. Instead, the pointwise methods score each passage based on the relevance with the query, which are less effective and stable. Therefore, the listwise methods are usually utilized to generate the ordered results, with sliding windows used to increase the efficiency, such as RankGPT. Considering the high cost of GPT API, we reproduce this method on LLaMA3-8b \cite{llama3modelcard}, which we refer to as RankLLaMA. Besides, to introduce the effectiveness of RankGPT, we select the DeBERTa-v3 as an alternative model that utilizes the knowledge-distilled dataset from GPT-4 to fine-tune DeBERTa \cite{He2020DeBERTaDB}.

\begin{figure}[t]
    \centering
    \includegraphics[width=0.8\linewidth]{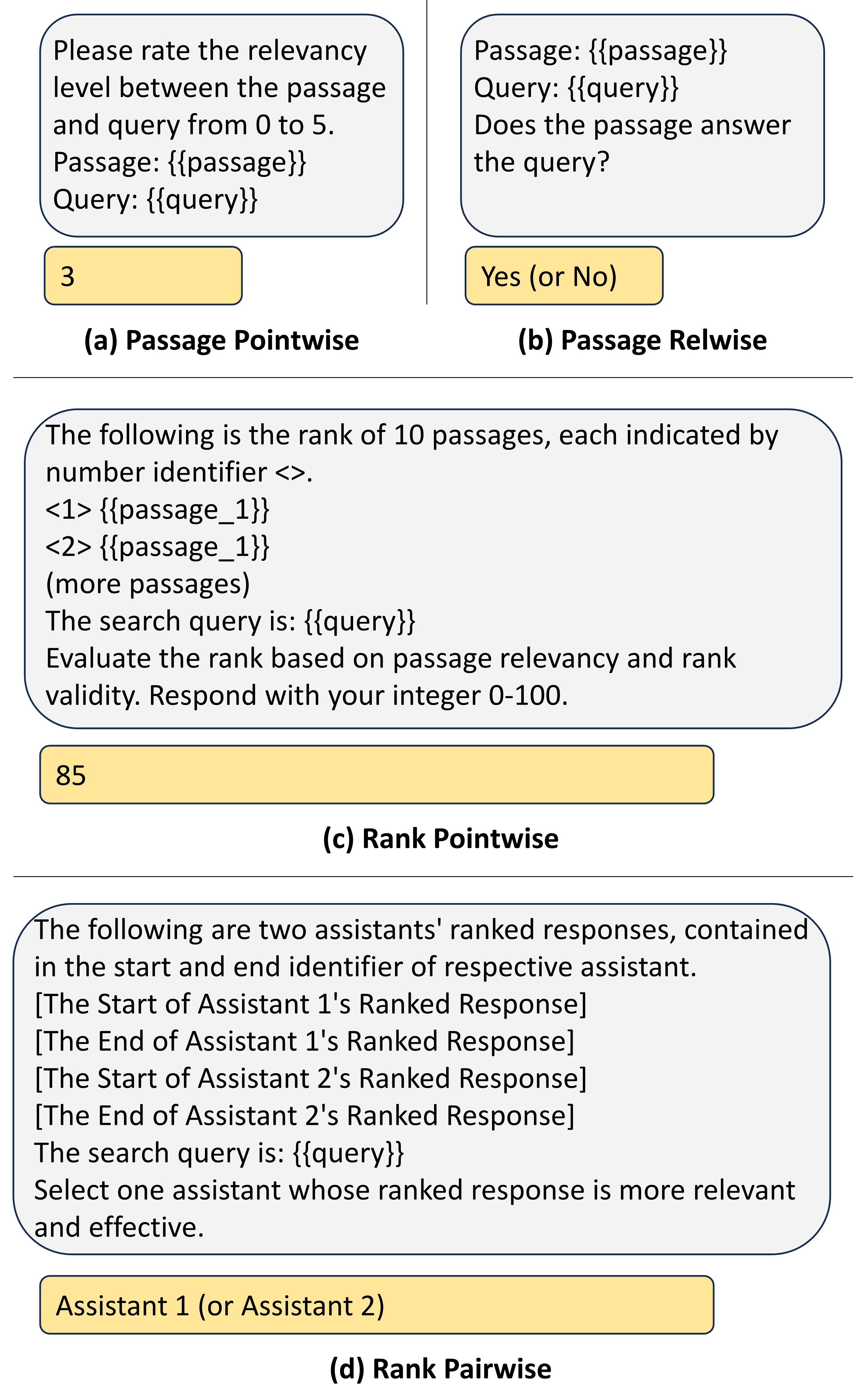}
    \caption{Four types of prompting methods for zero-shot rank-oriented automatic evaluation with LLMs. The gray and yellow blocks indicate the inputs and outputs of the model. (a) instructs LLMs to score the relevancy between the query and passage. (b) instructs LLMs to output relevance assessment. (c) rates an overall score for each rank. (d) directly selects the assistant with better ranks. Complete prompts in Appendix A.}
    \label{fig3}
\end{figure}

\subsection{Rank-Oriented Automatic Evaluation}

This task is designed to automatically evaluate the quality of ranked retrieval results based on their relevance to a given query. These methods focus on not only the individual relevance of passages but also the rationality of their ranking order, with the goal of picking out the most accurate and user-relevant results. In order to conduct reliable and comprehensive evaluation, we design four zero-shot prompting strategies without the reliance on annotated data. They can build an effective agent for the rank-oriented evaluation task with any open-source or black-box LLMs that don't require the training. Overall, these strategies can be divided into passage-based and rank-based methods, which mean a thought of rank-oriented evaluation respectively. These methods have their own strengths and weaknesses, and the detailed designs will be introduced as follows.



\subsubsection{Passage-based methods} 
These methods aim to simulate humans to evaluate the relevance of each retrieved passages and query, which can significantly reduce the context length and difficulty of the LLMs' evaluation. We first capture the passage collection without duplication and then propose some strategies to generate the relevance of each passage with the query. Specifically, we leverage the two types of prompting thoughts, pointwise and relwise. In the passage pointwise method, the LLMs are expected to annotate a score on the relevance within an internal, e.g. 0-5. This strategy emphasizes the subjectivity of LLMs and provides greater representation space for the relevancy. Instead, in the passage relwise method, we just utilize LLMs to make a judgment whether a passage and a query are relevant or not. The relevance is labeled as 0 or 1, which significantly reduces the evaluation difficulty. Finally, some popular machine evaluation metrics, such as NDCG, MAP and MRR, are used to calculate the overall score, which acts as the basis of selecting the best ranks.

\subsubsection{Rank-based methods} 
Because of the error transmission of passage-based methods, we design the rank-based methods to directly select the better rank based on LLMs. For the rank-oriented evaluation, in addition to considering relevance between the passages and query, the rationality of rank also needs to be taken into account. To overcome this challenge, we provide the detailed explanation of the evaluation requirements in the prompting instructions, as shown in the Appendix. Specifically, we also leverage the two types of prompting thoughts, pointwise and pairwise, to select the better rank. The rank will be inputted into LLMs by a unique identifier (e.g., <1>, <2>, etc.). Similar to the general evaluation with LLMs, the prompting strategies can enable LLMs to directly output the better rank they think without any intermediate steps. Note that the rank-oriented evaluation is challenging for LLMs, far beyond the text-oriented evaluation and passage-based prompting evaluation methods.

\begin{table*}[t]
    \centering
    \small
    \setlength\tabcolsep{2pt}
    \begin{tabular}{l cc | cccccccc | c }
         \toprule
         \textbf{Methods} & DL19 &  DL20 & Covid &  DBPedia & News &  NFCorpus & Robust04 &  SciFact & Signal1m &  Touche & BEIR(Avg) \\
         \midrule
         UPR (FLAN-T5-XL)
& 53.85 & 56.02 & 68.11 & 30.91 & 43.11 & 35.04 & 42.43 & 72.69 & 31.91 & 19.69 & 42.99
\\ 

InPars (monoT5-3B)
& - & 66.12 &  78.35 & - & - & - & - & - & - & - & -
\\

Promptagator++ (few-shot)
& - & - & 76.2 & 43.4 & - & 37.0 & - & 73.1 & - & 38.1 & -
\\

Cohere Rerank-v2
& 73.22 & 67.08 & 81.81 & 42.51 & 47.59 & 36.36 & 50.78 & 74.44 & 29.60 & 32.51 & 49.45
\\

ListT5-3B
& 71.80 & 69.10 & 84.70 & 46.20 & \textbf{53.20} & 37.70 & \textbf{57.80} & \textbf{77.00} & 33.80 & 33.60 & 53.00
\\

RankGPT (GPT3.5)
& 65.80 & 62.91  & 76.67 & 44.47 & 48.85 & 35.62 & 50.62 & 70.43 & 32.12 & 36.18 & 49.37
\\

RankGPT (GPT4)
& \textbf{75.59} & \textbf{70.56} & \textbf{85.51} & \textbf{47.12} & 52.89 & 38.47 & 57.55 & 74.95 & \textbf{34.40} & 38.57 & \textbf{53.68}
\\

\midrule
\textbf{Selected retrieval models} \\
\midrule

BM25
& 50.58 & 47.96 & 59.47 & 31.80 & 39.52 & 32.18 & 40.70 & 67.89 & 33.05 & \underline{\textbf{44.22}} & 43.60
\\

Contriever (110M)
& 67.49 & 66.6 & 67.87 & 41.39 & 41.67 & 32.97 & 46.50 & 66.85 & 27.83 & 24.77 & 43.73
\\

GTR-XXL (11B)
& 71.33 & 70.38 & 65.69 & 41.11 & 45.99 & 34.31 & 50.51 & 67.03 & 27.68 & 31.80 & 45.52
\\

MonoBERT (340M)
& 72.26 & 70.29 & 65.07 & 41.87 & 44.61 & 35.09 & 49.06 & 71.36 & 30.74 & 31.73 & 46.19
\\

MonoT5 (220M)
& 71.54 & 69.73 & 77.65 & 42.45 & 46.83 & 35.54 & 51.55 & 73.40 & 32.03 & 30.95 & 48.80
\\

MonoT5 (3B)
& 71.83 & 68.89 & 80.71 & 44.45 & 48.49 & \underline{\textbf{38.97}} & 56.71 & 76.57 & 32.55 & 32.41 & 51.36
\\

DeBERTa-3v (184M)
& 66.56 & 59.43 & 79.68 & 42.18 & 52.08 & 33.85 & 52.37 & 71.49 & \underline{33.41} & 38.01 & 50.38
\\

RankLLaMA (8B)
& 63.82 & 60.95 & 62.24 & 40.68 & 36.26 & 29.31 & 37.75 & 54.52 & 29.50 & 33.31 & 40.45
\\

\midrule
\textbf{Proposed Framework} \\
\midrule
DCRF (LLaMA3-8b)
& 74.29 & 70.59 & 82.24 & 46.68 & 48.83 & 37.38 & 57.02 & 76.54 & 31.28 & 33.6 & 51.70
\\

DCRF (LLaMA3-70b)
& 73.93 & 72.83 & \underline{\textbf{84.59}} & 46.80 & \underline{\textbf{52.65}} & \textbf{37.95} & 59.28 & \underline{\textbf{77.70}} & 31.74 & 32.73 & \underline{\textbf{52.93}}
\\

DCRF (Vicuna-7b)
& 74.27 & \underline{\textbf{74.07}} & 81.52 & 45.95 & 50.43 & 37.05 & 58.75 & 73.37 & \textbf{32.32} & 32.16 & 51.44
\\

DCRF (Vicuna-13b)
& \underline{\textbf{74.88}} & 74.02 & 81.70 & \underline{\textbf{47.63}} & 52.22 & 37.92 & \underline{\textbf{60.48}} & 76.31 & 31.05 & \textbf{33.72} & 52.63
\\

DCRF (ChatGLM-6b)
& 73.22 & 69.93 & 82.48 & 44.63 & 48.47 & 37.07 & 56.37 & 76.59 & 32.25 & 32.68 & 51.32
\\

DCRF (ChatGLM2-6b)
& 74.38 & 70.03 & 80.14 & 45.51 & 46.77 & 36.72 & 57.10 & 75.69 & 32.71 & 33.00 & 50.96
\\

\bottomrule
\end{tabular}
\caption{Results (nDCG@10) on TREC and BEIR. All models except InPars, Promptagator++ and RankGPT (GPT4) re-rank the same BM25 top-100 passages. The best performances of existing models and DCRF are marked bold, and the better effects between single models and the collaborative framework are underlined.}
\label{tab:benchmarks}
\end{table*}

\section{Experiments}

\subsection{Experimental Settings}

Following the previous influential retrieval work, RankGPT \cite{Sun2023IsCG}, we utilize the benchmark datasets including TREC-DL19, TREC-DL20 \cite{Craswell2021OverviewOT} and BEIR \cite{Thakur2021BEIRAH}. In BEIR, we choose eight tasks to evaluate the models that are consistent with RankGPT. To verify the effectiveness of DCRF on different LLMs, we choose the open-source LLaMA3-8b, LLaMA3-70b \cite{llama3modelcard}, Vicuna-7b-v1.5, Vicuna-13b-v1.3 \cite{Zheng2023JudgingLW}, ChatGLM-6b and ChatGLM2-6b \cite{Zeng2024ChatGLMAF}, which contain different training data, model architectures and parameter quantities. Besides, considering the high cost of black-box LLM APIs, we utilize the GPT-3.5, GPT-4 and CLAUDE-3.5 as evaluation agents on several representative datasets.

\subsection{Implement Details}

The first-stage retrieval utilizes the BM25 algorithm with Pyserini \footnote{https://github.com/castorini/pyserini} to recall the top-100 passages. In the second-stage rerank, we reproduce the inference process based on the trained model parameters on the Hugging Face \footnote{https://huggingface.co/}. In the third-stage evaluation, in order to ensure the reproducibility, we set the do\_sample to False and temperature to 0 for all LLMs. The entire experiments can be conducted on an 8-card 3090 server.

\subsection{Effectiveness and Potential Analysis}

In theory, our proposed DCRF can integrate any retrieval model and achieve the best performance of IR system at present. In order to test the full potential of this framework, we conduct a series of analysis experiments and find that the proposed framework has excellent performance boundary.

First of all, following previous works, we use the NDCG@10 as the criterion to assess the performance of retrieval models for each query and corpus and count the frequency as shown in Figure \ref{fig1}. On the whole, the performance of all retrieval models is relatively balanced and no model is able to achieve good frequency across more than three datasets. This indicates that no model can currently adapt to all queries and corpora, verifying our assumption. In the other word, all the existing retrieval methods have their own preference scenarios. For example, the rates of BM25 are less than 10\% in the DL19, DL20 and Covid, while they are more than 30\% in the NFCorpus, Scifact and Touche. This is also the evidence supporting the conclusion that BM25 has better performance for professional domains and long documents. Instead, the model with the best average performance, MonoT5-3b, generally has the frequency between 10\% and 20\%, demonstrating its universality and generalization.

\begin{figure}[t]
    \centering
    \includegraphics[width=0.97\linewidth]{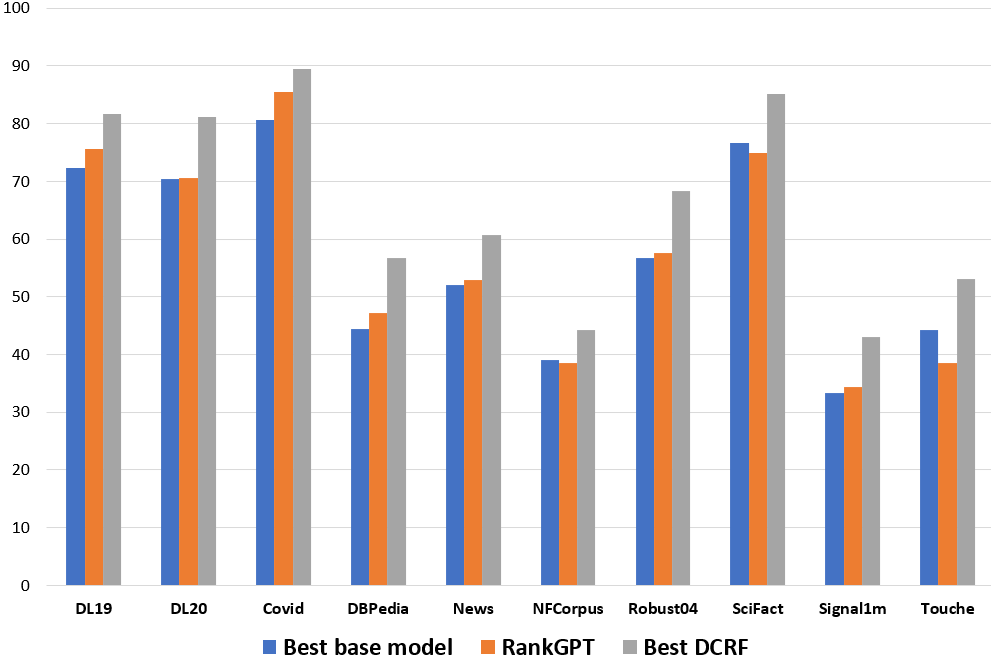}
    \caption{Comparison for ideal scenarios on TREC and BEIR. the Best base model in the legend indicates the best performance among all selected retrieval models. Best DCRF indicates the performance of the framework with NDCG@10 as evaluation, which relies on the labels annotated by humans.}
    \label{fig:analisys}
\end{figure}

Next, in order to further analyse the potential of DCRF, we simulate the optimal performance by utilizing supervised metrics as the evaluator. As shown in Figure \ref{fig:analisys}, best DCRF based on selected efficient retrieval models outperforms the start-of-the-art model. Typically, there are more than 10\% improvements on the general-domain dataset, DL20 and the scientific-domain dataset, SciFact. These fully demonstrate the potential of the DCRF and the value of this idea. However, the NDCG evaluation depends on the annotated relevance score, which cannot be directly used in the applications. Therefore, we focus on unsupervised rank-oriented evaluation methods to help DCRF approach its upper bound of performance.

\subsection{Main Results}

On the benchmarks, we compare the DCRF with selected retrieval models and other state-of-the-art retrieval methods. They include: UPR \cite{Sachan2022ImprovingPR}, InPairs \cite{Bonifacio2022InParsDA}, Promptagator++ \cite{Dai2022PromptagatorFD}, Cohere Rerank \footnote{https://txt.cohere.com/rerank/}, ListT5 \cite{Yoon2024ListT5LR} and RankGPT \cite{Sun2023IsCG}. Then we utilize LLaMA3, Vicuna-v1.5 and ChatGLM to act as the zero-shot evaluation agents. To adapt to different queries and corpora, we choose the most effective prompting strategy for each dataset on the specific LLMs. The experimental results are shown in Table \ref{tab:benchmarks}.

On the whole, the proposed models outperform the selected retrieval models, demonstrating the effectiveness of the prompting evaluation strategies. Besides, the DCRF (LLaMA3-70b) can achieve better performance and faster inference compared to existing models based on LLMs, such as RankGPT (GPT3.5), indicating the outstanding performance and efficiency of the proposed framework. Surprisingly, the chosen open-source LLMs, as evaluation agents without training, can achieve comparable performance to the strong black-box LLMs, RankGPT (GPT4) and trained language model, ListT5, which also provides the momentum for further research on instruction tuning in rank-oriented automatic evaluation. 

Moreover, we also find some other interesting conclusions from experimental results: (1) DCRF across various LLMs can outperform all the selected retrieval models. This illustrates the LLMs have the capacity to become an effective rank-oriented evaluation agent. (2) LLMs with more parameters have better performance. This reflects the effectiveness of the proposed prompting strategies is constrained by model capability, especially in terms of the unseen rank-oriented evaluation ability. It's a motivation to further instruction-tune a better rank-oriented evaluation model. (3) Datasets with richer knowledge and features can better highlight the differences among models, such as DBpedia-Entity. In contrast, there are only minuscule differences of DCRF on general tasks like DL19.

\begin{table}[t]
    \centering
    \small
    \setlength\tabcolsep{5pt}
    \begin{tabular}{l cccc}
        \toprule     
        \textbf{} & DL19 & Covid & DBPedia & SciFact \\
         \midrule
         \textbf{Supervised Metrics} \\
         \midrule
        NDCG@10 & 81.61 & 89.47 & 56.76 & 85.18 \\
        MAP@10 & 76.65 & 87.28 & 55.51 & 85.18 \\
        MRR@10 & 72.38 & 71.29 & 48.63 & 84.16 \\
         \midrule
         \textbf{LLaMA3-8b} \\
         \midrule
        Passage-Pointwise & 72.65 & \textbf{82.24} & \textbf{46.68} & 72.06 \\
        Passage-Relwise & \textbf{74.29} & 78.62 & 45.51 & 75.39 \\
        Rank-Pointwise & 72.83 & 80.71 & 44.43 & \textbf{76.54} \\
        Rank-Pairwise & 73.07 & 80.95 & 44.79 & 76.37 \\
        \midrule
         \textbf{LLaMA3-70b}
          \\
         \midrule
        Passage-Pointwise & \textbf{73.93} & \textbf{84.59} & 46.70 & \textbf{77.70} \\
        Passage-Relwise & 73.48 & 81.58 & 46.17 & 75.23 \\
        Rank-Pointwise & 72.81 & 80.71 & 44.33 & 76.59 \\
        Rank-Pairwise & 73.57 & 79.81 & \textbf{46.80} & 76.15 \\
         \bottomrule
    \end{tabular}
    \caption{Compare different prompting methods and supervised IR metrics. Best performing prompting methods under the same LLMs are marked bold. The typical datasets are selected to show and see Appendix B for the complete results with all benchmarks and LLMs.}
    \label{tab:effect_of_instructions}
\end{table}

\subsection{Ablation Studies}

\subsubsection{Effect of prompting strategies}

In order to compare the performance of four prompting strategies, we repeat the experiments with all prompting methods on all datasets and LLMs. Meanwhile, to compare with the ideal effect of DCRF, we introduce the supervised metrics as the evaluator which needs annotated references. The typical results are shown in Table \ref{tab:effect_of_instructions}. On the whole, the performance of passage-based evaluation methods is better than that of rank-based evaluation methods on both LLaMA3-8b and LLaMA3-70b. This indicates that existing LLMs are hard to directly process the complex rank-based evaluation at present, which illustrates that it's still a challenge to arouse LLMs' capacity on the rank-oriented automatic evaluation task. Instead, more specific passage-based methods are more suitable for current LLMs. Next, the results indicate that the LLM with more parameters prefers pointwise than relwise, which is consistent with the examples in Figure \ref{fig3}, where passage-pointwise is a more difficult instruction than passage-relwise. Therefore, when LLMs with a sufficient number of parameters are utilized, passage-pointwise is the optimal instructional prompting method. However, there is also large gap between supervised metrics and unsupervised automatic evaluation methods, indicating the potential for further research.

\subsubsection{Effect of LLMs}

\begin{table}[t]
    \centering
    \small
    \begin{tabular}{l cccc}
        \toprule     
        \textbf{Models} & DL19 & Covid & News & Touche \\
         \midrule
         \textbf{Open-Source LLMs} \\
         \midrule
        LLaMA3-8b & 73.97 & 80.71 & 47.55 & 31.54\\
        LLaMA3-70b & 73.93 & 84.59 & 52.65 & 32.32\\
        Vicuna-v1.5-7b & 73.73 & 80.69 & 49.23 & 29.84 \\
        Vicuna-v1.5-13b & 74.88 & 72.79 & 52.22 & 31.83 \\
        Chatglm-6b & 66.70 & 65.38 & 42.89 & 27.19 \\
        Chatglm2-6b & 64.65 & 71.07 & 44.05 & 33.00 \\
         \midrule
         \textbf{Black-Box LLMs} \\
         \midrule
        GPT-3.5 & 73.00 & 81.84 & 50.38 & 33.48 \\
        GPT-4 & 74.48 & 83.13 & 49.94 & 31.91 \\
        CLAUDE-3.5 & 74.68 & 83.30 & 51.44 & 34.83 \\
         \bottomrule
    \end{tabular}
    \caption{Compare with various open-source and black-box LLMs as evaluation agents. The passage-pointwise is selected as the instructional method, which is the most suitable for LLMs with more parameters.}
    \label{tab:effect_of_llms}
\end{table}

Besides open-source LLMs, the proposed DCRF can directly expand to any black-box LLMs and we conduct experiments to study the effect of different LLMs. To reduce the cost of API calling, we select several typical datasets, including DL19, Covid, TREC-News and Touche. As shown in Table \ref{tab:effect_of_llms}, it is clear that models with more parameters have overall better performance, such as LLaMA3-70b. However, there is no significant difference in performance between open-source and black-box LLMs, such as LLaMA3-70b and GPT-4. We think the reason is that for the unseen and difficult task, rank-oriented automatic evaluation, our designed prompting strategies are effective enough to reach the upper bound of existing LLMs's capacity without training. The generated results illustrate the scores rated by LLaMA3-70b and GPT-4 are both reasonable, verifying our thought. Besides, it also indicates that knowledge distillation may not have significant impact. Instead, the instruction tuning and in-context learning could be the key to further enhancing the LLMs' capacity on the rank-oriented automatic evaluation. The total cost of black-box LLMs amounts to about \$77.5.

\subsection{Analysis}

There are two additional important problems in the DCRF: (1) what is the inference efficiency loss when using an additional module compared to the traditional IR framework. (2) does the bias of LLMs as evaluators exist in the DCRF, as mentioned by previous works \cite{Wang2023LargeLM, Stureborg2024LargeLM}.

{\renewcommand{\arraystretch}{1.1} 
\begin{table}[t]
    \small
    \centering
    \vspace{-1em}
    \resizebox{0.44\textwidth}{!}{
    \begin{tabular}{l|c}
        \toprule
        Methods & Inference cost \\
        \midrule
        Passage-based & $L_{p} * T_{LLM}$ \\
        Rank-based  &  \\
        \hspace{1em} -pointwise & $k * L_{p} * T_{LLM}$ \\
        \hspace{1em} -pairwise & $N_{ranks} * k * L_{p} * T_{LLM}$ \\
        \midrule
        RankGPT & $ N_{step} * S_{windows} * L_{p} * T_{LLM}$ \\
        ListT5 & $O(N + k * logN) * L_{p}  * T_{LM}$ \\
        \bottomrule
    \end{tabular}
    }
    \caption{The quantitative comparison of inference cost between DCRF with prompting strategies and other models. $L_{p}$ means the average length of the passages and $k$ means the number of selected passages in a rank, generally set to 10. $N_{ranks}$ indicates the number of selected retrieval models, set to 8 in our experiments. $N_{step}$ and $S_{windows}$ indicate the step and window size of sliding windows algorithm respectively, whose product is usually 200. The $N$ in the ListT5 means the number of candidates that is usually 100.}
    \label{tab:quantitative_analysis}
\end{table}
}

\subsubsection{Quantitative analysis of inference cost}

In order to ensure the practicability, we need to evaluate the inference cost of DCRF, as shown in Table \ref{tab:quantitative_analysis}. Because all evaluation processes can be computed in parallel, the inference cost mainly depends on the context length and reasoning frequency. The results show that the passage-based strategies have lower inference cost than the rank-based strategies and the rank-based pairwise method will spend the longest time in the practical applications. Considering the little time required for LLMs to perform one inference, DCRF doesn't significantly increase the cost compared to traditional IR framework. Besides, it has better inference efficiency than RankGPT and ListT5, because of the distributed designs and parallel computing. Note that the time of selected retrieval models is ignored, which is minimal due to pre-processing and small models compared to the evaluator.

\begin{figure}[t]
    \centering
    \includegraphics[width=\linewidth]{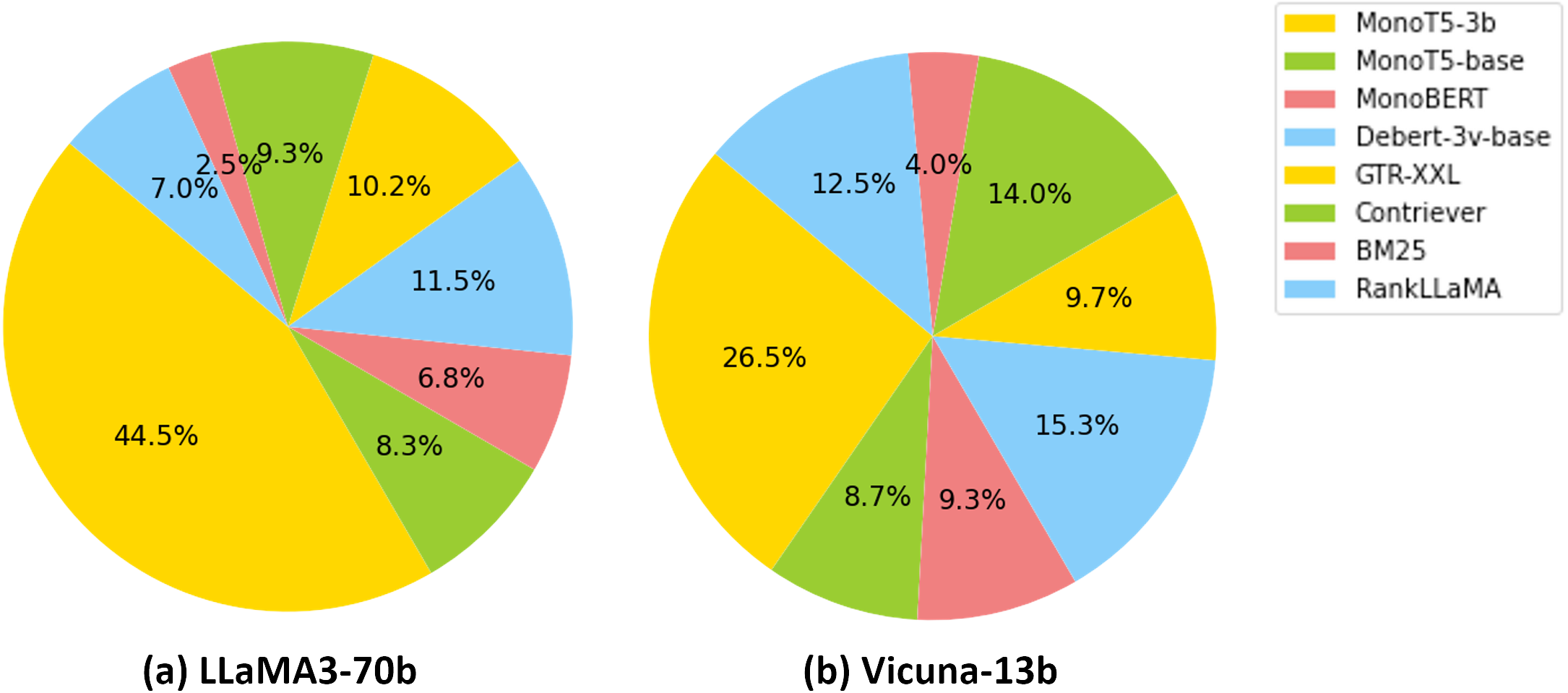}
    \caption{Statistics of selected retrieval models as the best model on Dbpedia-Entity. The order of retrieval models in the legend and evaluation is consistent. The passage-pointwise is selected as the prompting method.}
    \label{fig:bias_of_llms}
\end{figure}

\subsubsection{Bias of LLMs as evaluators}

\citet{Wang2023LargeLM} demonstrate that there is a systematic positional bias in evaluation with LLMs, which could compromise the fairness of evaluation. In order to study whether the position of retrieval models also affects the performance of the evaluators, we record the rate of selected retrieval models as the best model on Dbpedia-Entity, as shown in Figure \ref{fig:bias_of_llms}. It can be found there is no tendency for models located earlier in the order to be more frequently selected. For example, DeBERTa-v3-base and Contriever account for a larger proportion compared to MonoT5-base and MonoBERT. This supports the validity of the rank-oriented evaluation, because models with different architectures have an equitable impact on the final performance. Besides, this also suggests the positional bias is minimal for proposed rank-oriented evaluation strategies, contributing to a fair integration of various retrieval models.

\section{Conclusion}

In this paper, we propose a novel Distributed Collaborative Retrieval Framework (DCRF) to integrate various retrieval models into a unified system, which aims to achieve the optimal performance across all the queries and corpora. The preliminary analysis experiments illustrate the excellent performance boundary and research values of the proposed framework. Next, we design four effective prompting strategies with LLMs to perform the rank-oriented automatic evaluation without reliance of labeled data. Extensive experiments on 10 benchmarks and 6 LLMs demonstrate the effectiveness and efficiency of DCRF and our prompting strategies. Finally, this framework is highly flexible and extensible, enabling the integration of any new model to accommodate emerging scenarios.

\section{Limitations}

DCRF is the first distributed collaborative framework to integrate various retrieval models as far as we know, although it still exhibits certain limitations that warrant future improvements. (1) As the key part of the framework, the rank-oriented evaluator can be further optimized by instruction tuning or in-context learning, in order to approach the DCRF's performance boundary illustrated by the preliminary analysis experiments. (2) The more domain-adaptive retrieval models haven't been selected due to unpublished model parameters. We can train retrieval models on specific domain corpus and integrate them to the DCRF, in order to improve the its domain generalization ability. (3) We haven't further analyse the data features of the benchmarks and categorize them into specific defined scenarios, which can reveal the mechanistic reasons for models' preferences. 





\bibliography{custom}

\begin{thebibliography}{34}
\providecommand{\natexlab}[1]{#1}

\bibitem[{AI@Meta(2024)}]{llama3modelcard}
AI@Meta. 2024.
\newblock \href {https://github.com/metallama/llama3/blob/main/MODEL_CARD.md}
  {Llama 3 model card}.

\bibitem[{Bonifacio et~al.(2022)Bonifacio, Abonizio, Fadaee, and
  Nogueira}]{Bonifacio2022InParsDA}
Luiz~Henrique Bonifacio, Hugo Abonizio, Marzieh Fadaee, and Rodrigo Nogueira.
  2022.
\newblock \href {https://api.semanticscholar.org/CorpusID:246705967} {Inpars:
  Data augmentation for information retrieval using large language models}.
\newblock \emph{ArXiv}, abs/2202.05144.

\bibitem[{Craswell et~al.(2021)Craswell, Mitra, Yilmaz, Campos, and
  Voorhees}]{Craswell2021OverviewOT}
Nick Craswell, Bhaskar Mitra, Emine Yilmaz, Daniel~Fernando Campos, and
  Ellen~M. Voorhees. 2021.
\newblock \href {https://api.semanticscholar.org/CorpusID:212737158} {Overview
  of the trec 2020 deep learning track}.
\newblock \emph{ArXiv}, abs/2102.07662.

\bibitem[{Dai et~al.(2022)Dai, Zhao, Ma, Luan, Ni, Lu, Bakalov, Guu, Hall, and
  Chang}]{Dai2022PromptagatorFD}
Zhuyun Dai, Vincent Zhao, Ji~Ma, Yi~Luan, Jianmo Ni, Jing Lu, Anton Bakalov,
  Kelvin Guu, Keith~B. Hall, and Ming-Wei Chang. 2022.
\newblock \href {https://api.semanticscholar.org/CorpusID:252519173}
  {Promptagator: Few-shot dense retrieval from 8 examples}.
\newblock \emph{ArXiv}, abs/2209.11755.

\bibitem[{Ferraretto et~al.(2023)Ferraretto, Laitz, de~Alencar~Lotufo, and
  Nogueira}]{Ferraretto2023ExaRankerEN}
Fernando Ferraretto, Thiago Laitz, Roberto de~Alencar~Lotufo, and Rodrigo
  Nogueira. 2023.
\newblock \href {https://api.semanticscholar.org/CorpusID:256231455}
  {Exaranker: Explanation-augmented neural ranker}.
\newblock \emph{ArXiv}, abs/2301.10521.

\bibitem[{Fu et~al.(2023)Fu, Ng, Jiang, and Liu}]{fu2023gptscore}
Jinlan Fu, See-Kiong Ng, Zhengbao Jiang, and Pengfei Liu. 2023.
\newblock Gptscore: Evaluate as you desire.
\newblock \emph{arXiv preprint arXiv:2302.04166}.

\bibitem[{He et~al.(2020)He, Liu, Gao, and Chen}]{He2020DeBERTaDB}
Pengcheng He, Xiaodong Liu, Jianfeng Gao, and Weizhu Chen. 2020.
\newblock \href {https://api.semanticscholar.org/CorpusID:219531210} {Deberta:
  Decoding-enhanced bert with disentangled attention}.
\newblock \emph{ArXiv}, abs/2006.03654.

\bibitem[{Izacard et~al.(2021)Izacard, Caron, Hosseini, Riedel, Bojanowski,
  Joulin, and Grave}]{Izacard2021UnsupervisedDI}
Gautier Izacard, Mathilde Caron, Lucas Hosseini, Sebastian Riedel, Piotr
  Bojanowski, Armand Joulin, and Edouard Grave. 2021.
\newblock \href {https://api.semanticscholar.org/CorpusID:249097975}
  {Unsupervised dense information retrieval with contrastive learning}.
\newblock \emph{Trans. Mach. Learn. Res.}, 2022.

\bibitem[{Karpukhin et~al.(2020)Karpukhin, Oğuz, Min, Lewis, Wu, Edunov, Chen,
  and tau Yih}]{Karpukhin2020DensePR}
Vladimir Karpukhin, Barlas Oğuz, Sewon Min, Patrick Lewis, Ledell~Yu Wu,
  Sergey Edunov, Danqi Chen, and Wen tau Yih. 2020.
\newblock \href {https://api.semanticscholar.org/CorpusID:215737187} {Dense
  passage retrieval for open-domain question answering}.
\newblock \emph{ArXiv}, abs/2004.04906.

\bibitem[{Khattab and Zaharia(2020)}]{Khattab2020ColBERTEA}
O.~Khattab and Matei~A. Zaharia. 2020.
\newblock \href {https://api.semanticscholar.org/CorpusID:216553223} {Colbert:
  Efficient and effective passage search via contextualized late interaction
  over bert}.
\newblock \emph{Proceedings of the 43rd International ACM SIGIR Conference on
  Research and Development in Information Retrieval}.

\bibitem[{Kocmi and Federmann(2023)}]{Kocmi2023LargeLM}
Tom Kocmi and Christian Federmann. 2023.
\newblock \href {https://api.semanticscholar.org/CorpusID:257232490} {Large
  language models are state-of-the-art evaluators of translation quality}.
\newblock \emph{ArXiv}.

\bibitem[{Luo et~al.(2023)Luo, Xie, and Ananiadou}]{Luo2023ChatGPTAA}
Zheheng Luo, Qianqian Xie, and Sophia Ananiadou. 2023.
\newblock \href {https://api.semanticscholar.org/CorpusID:258108400} {Chatgpt
  as a factual inconsistency evaluator for text summarization}.
\newblock \emph{ArXiv}.

\bibitem[{Ni et~al.(2021)Ni, Qu, Lu, Dai, Abrego, Ma, Zhao, Luan, Hall, Chang,
  and Yang}]{Ni2021LargeDE}
Jianmo Ni, Chen Qu, Jing Lu, Zhuyun Dai, Gustavo~Hern{\'a}ndez Abrego, Ji~Ma,
  Vincent Zhao, Yi~Luan, Keith~B. Hall, Ming-Wei Chang, and Yinfei Yang. 2021.
\newblock \href {https://api.semanticscholar.org/CorpusID:245144556} {Large
  dual encoders are generalizable retrievers}.
\newblock \emph{ArXiv}, abs/2112.07899.

\bibitem[{Nogueira and Cho(2019)}]{Nogueira2019PassageRW}
Rodrigo Nogueira and Kyunghyun Cho. 2019.
\newblock \href {https://api.semanticscholar.org/CorpusID:58004692} {Passage
  re-ranking with bert}.
\newblock \emph{ArXiv}, abs/1901.04085.

\bibitem[{Nogueira et~al.(2020)Nogueira, Jiang, Pradeep, and
  Lin}]{Nogueira2020DocumentRW}
Rodrigo Nogueira, Zhiying Jiang, Ronak Pradeep, and Jimmy~J. Lin. 2020.
\newblock \href {https://api.semanticscholar.org/CorpusID:212725651} {Document
  ranking with a pretrained sequence-to-sequence model}.
\newblock \emph{ACL}.

\bibitem[{Panickssery et~al.(2024)Panickssery, Bowman, and
  Feng}]{Panickssery2024LLMER}
Arjun Panickssery, Samuel~R. Bowman, and Shi Feng. 2024.
\newblock \href {https://api.semanticscholar.org/CorpusID:269293311} {Llm
  evaluators recognize and favor their own generations}.
\newblock \emph{ArXiv}, abs/2404.13076.

\bibitem[{Pradeep et~al.(2023)Pradeep, Sharifymoghaddam, and
  Lin}]{Pradeep2023RankZephyrEA}
Ronak Pradeep, Sahel Sharifymoghaddam, and Jimmy~J. Lin. 2023.
\newblock \href {https://api.semanticscholar.org/CorpusID:265659387}
  {Rankzephyr: Effective and robust zero-shot listwise reranking is a breeze!}
\newblock \emph{ArXiv}, abs/2312.02724.

\bibitem[{Rei et~al.(2021)Rei, Farinha, Zerva, van Stigt, Stewart, Ramos,
  Glushkova, Martins, and Lavie}]{rei2021references}
Ricardo Rei, Ana~C Farinha, Chrysoula Zerva, Daan van Stigt, Craig Stewart,
  Pedro Ramos, Taisiya Glushkova, Andr{\'e}~FT Martins, and Alon Lavie. 2021.
\newblock Are references really needed? unbabel-ist 2021 submission for the
  metrics shared task.
\newblock In \emph{Proceedings of the Sixth Conference on Machine Translation},
  pages 1030--1040.

\bibitem[{Robertson and Zaragoza(2009)}]{Robertson2009ThePR}
Stephen~E. Robertson and Hugo Zaragoza. 2009.
\newblock \href {https://api.semanticscholar.org/CorpusID:207178704} {The
  probabilistic relevance framework: Bm25 and beyond}.
\newblock \emph{Found. Trends Inf. Retr.}, 3:333--389.

\bibitem[{Rosa et~al.(2022)Rosa, Bonifacio, Jeronymo, Abonizio, Fadaee,
  de~Alencar~Lotufo, and Nogueira}]{Rosa2022InDO}
Guilherme~Moraes Rosa, Luiz~Henrique Bonifacio, Vitor Jeronymo, Hugo Abonizio,
  Marzieh Fadaee, Roberto de~Alencar~Lotufo, and Rodrigo Nogueira. 2022.
\newblock \href {https://api.semanticscholar.org/CorpusID:254564419} {In
  defense of cross-encoders for zero-shot retrieval}.
\newblock \emph{ArXiv}, abs/2212.06121.

\bibitem[{Sachan et~al.(2022)Sachan, Lewis, Joshi, Aghajanyan, tau Yih, Pineau,
  and Zettlemoyer}]{Sachan2022ImprovingPR}
Devendra~Singh Sachan, Mike Lewis, Mandar Joshi, Armen Aghajanyan, Wen tau Yih,
  Jo{\"e}lle Pineau, and Luke Zettlemoyer. 2022.
\newblock \href {https://api.semanticscholar.org/CorpusID:248218489} {Improving
  passage retrieval with zero-shot question generation}.
\newblock \emph{Conference on Empirical Methods in Natural Language
  Processing}.

\bibitem[{Scialom et~al.(2019)Scialom, Lamprier, Piwowarski, and
  Staiano}]{scialom2019answers}
Thomas Scialom, Sylvain Lamprier, Benjamin Piwowarski, and Jacopo Staiano.
  2019.
\newblock Answers unite! unsupervised metrics for reinforced summarization
  models.
\newblock \emph{arXiv preprint arXiv:1909.01610}.

\bibitem[{Shen et~al.(2023)Shen, Cheng, You, and Bing}]{Shen2023LargeLM}
Chenhui Shen, Liying Cheng, Yang You, and Lidong Bing. 2023.
\newblock \href {https://api.semanticscholar.org/CorpusID:258833685} {Large
  language models are not yet human-level evaluators for abstractive
  summarization}.
\newblock \emph{ArXiv}.

\bibitem[{Stureborg et~al.(2024)Stureborg, Alikaniotis, and
  Suhara}]{Stureborg2024LargeLM}
Rickard Stureborg, Dimitris Alikaniotis, and Yoshi Suhara. 2024.
\newblock \href {https://api.semanticscholar.org/CorpusID:269588132} {Large
  language models are inconsistent and biased evaluators}.
\newblock \emph{ArXiv}, abs/2405.01724.

\bibitem[{Sun et~al.(2023)Sun, Yan, Ma, Ren, Yin, and Ren}]{Sun2023IsCG}
Weiwei Sun, Lingyong Yan, Xinyu Ma, Pengjie Ren, Dawei Yin, and Zhaochun Ren.
  2023.
\newblock \href {https://api.semanticscholar.org/CorpusID:258212638} {Is
  chatgpt good at search? investigating large language models as re-ranking
  agent}.
\newblock \emph{ArXiv}, abs/2304.09542.

\bibitem[{Thakur et~al.(2021)Thakur, Reimers, Ruckl'e, Srivastava, and
  Gurevych}]{Thakur2021BEIRAH}
Nandan Thakur, Nils Reimers, Andreas Ruckl'e, Abhishek Srivastava, and Iryna
  Gurevych. 2021.
\newblock \href {https://api.semanticscholar.org/CorpusID:233296016} {Beir: A
  heterogenous benchmark for zero-shot evaluation of information retrieval
  models}.
\newblock \emph{ArXiv}, abs/2104.08663.

\bibitem[{Vasilyev et~al.(2020)Vasilyev, Dharnidharka, and
  Bohannon}]{vasilyev2020fill}
Oleg Vasilyev, Vedant Dharnidharka, and John Bohannon. 2020.
\newblock Fill in the blanc: Human-free quality estimation of document
  summaries.
\newblock \emph{arXiv preprint arXiv:2002.09836}.

\bibitem[{Wang et~al.(2024)Wang, Huang, Tu, Wang, Huang, Laskar, and
  Bhuiyan}]{Wang2024UtilizingBF}
Jiajia Wang, Jimmy~X. Huang, Xinhui Tu, Junmei Wang, Angela~J. Huang,
  Md~Tahmid~Rahman Laskar, and Amran Bhuiyan. 2024.
\newblock \href {https://api.semanticscholar.org/CorpusID:267707681} {Utilizing
  bert for information retrieval: Survey, applications, resources, and
  challenges}.
\newblock \emph{ACM Computing Surveys}, 56:1 -- 33.

\bibitem[{Wang et~al.(2023)Wang, Li, Chen, Zhu, Lin, Cao, Liu, Liu, and
  Sui}]{Wang2023LargeLM}
Peiyi Wang, Lei Li, Liang Chen, Dawei Zhu, Binghuai Lin, Yunbo Cao, Qi~Liu,
  Tianyu Liu, and Zhifang Sui. 2023.
\newblock \href {https://api.semanticscholar.org/CorpusID:258960339} {Large
  language models are not fair evaluators}.
\newblock \emph{ArXiv}, abs/2305.17926.

\bibitem[{Yoon et~al.(2024)Yoon, Choi, Kim, Kim, Yun, and won
  Hwang}]{Yoon2024ListT5LR}
Soyoung Yoon, Eunbi Choi, Jiyeon Kim, Yireun Kim, Hyeongu Yun, and Seung won
  Hwang. 2024.
\newblock \href {https://api.semanticscholar.org/CorpusID:267938301} {Listt5:
  Listwise reranking with fusion-in-decoder improves zero-shot retrieval}.
\newblock \emph{ArXiv}, abs/2402.15838.

\bibitem[{Zeng et~al.(2024)Zeng, Xu, Wang, Zhang, Yin, Rojas, Feng, Zhao, Lai,
  Yu, Wang, Sun, Zhang, Cheng, Gui, Tang, Zhang, Li, Zhao, Wu, Zhong, yue Liu,
  Huang, Zhang, Zheng, Lu, Duan, Zhang, Cao, Yang, Tam, Zhao, Liu, Xia, Zhang,
  Gu, Lv, Liu, Liu, Yang, Song, Zhang, An, Xu, Niu, Yang, Li, Bai, Dong, Qi,
  Wang, Yang, Du, Hou, and Wang}]{Zeng2024ChatGLMAF}
Team Glm~Aohan Zeng, Bin Xu, Bowen Wang, Chenhui Zhang, Da~Yin, Diego Rojas,
  Guanyu Feng, Hanlin Zhao, Hanyu Lai, Hao Yu, Hongning Wang, Jiadai Sun,
  Jiajie Zhang, Jiale Cheng, Jiayi Gui, Jie Tang, Jing Zhang, Juanzi Li, Lei
  Zhao, Lindong Wu, Lucen Zhong, Ming yue Liu, Minlie Huang, Peng Zhang, Qinkai
  Zheng, Rui Lu, Shuaiqi Duan, Shudan Zhang, Shulin Cao, Shuxun Yang, Weng~Lam
  Tam, Wenyi Zhao, Xiao Liu, Xiaoyu Xia, Xiaohan Zhang, Xiaotao Gu, Xin Lv,
  Xinghan Liu, Xinyi Liu, Xinyue Yang, Xixuan Song, Xunkai Zhang, Yi~An, Yifan
  Xu, Yilin Niu, Yuantao Yang, Yueyan Li, Yushi Bai, Yuxiao Dong, Zehan Qi,
  Zhaoyu Wang, Zhenyi Yang, Zhengxiao Du, Zhen-Ping Hou, and Zihan Wang. 2024.
\newblock \href {https://api.semanticscholar.org/CorpusID:270562306} {Chatglm:
  A family of large language models from glm-130b to glm-4 all tools}.
\newblock \emph{ArXiv}, abs/2406.12793.

\bibitem[{Zhao et~al.(2022)Zhao, Liu, Ren, and rong Wen}]{Zhao2022DenseTR}
Wayne~Xin Zhao, Jing Liu, Ruiyang Ren, and Ji~rong Wen. 2022.
\newblock \href {https://api.semanticscholar.org/CorpusID:254044526} {Dense
  text retrieval based on pretrained language models: A survey}.
\newblock \emph{ACM Transactions on Information Systems}, 42:1 -- 60.

\bibitem[{Zheng et~al.(2023)Zheng, Chiang, Sheng, Zhuang, Wu, Zhuang, Lin, Li,
  Li, Xing, Zhang, Gonzalez, and Stoica}]{Zheng2023JudgingLW}
Lianmin Zheng, Wei-Lin Chiang, Ying Sheng, Siyuan Zhuang, Zhanghao Wu, Yonghao
  Zhuang, Zi~Lin, Zhuohan Li, Dacheng Li, Eric~P. Xing, Haotong Zhang, Joseph
  Gonzalez, and Ion Stoica. 2023.
\newblock \href {https://api.semanticscholar.org/CorpusID:259129398} {Judging
  llm-as-a-judge with mt-bench and chatbot arena}.
\newblock \emph{ArXiv}, abs/2306.05685.

\bibitem[{Zhu et~al.(2023)Zhu, Yuan, Wang, Liu, Liu, Deng, Dou, and rong
  Wen}]{Zhu2023LargeLM}
Yutao Zhu, Huaying Yuan, Shuting Wang, Jiongnan Liu, Wenhan Liu, Chenlong Deng,
  Zhicheng Dou, and Ji~rong Wen. 2023.
\newblock \href {https://api.semanticscholar.org/CorpusID:260887838} {Large
  language models for information retrieval: A survey}.
\newblock \emph{ArXiv}, abs/2308.07107.

\end{thebibliography}

\onecolumn
\clearpage
\appendix

\section{Instructional Prompts}

\subsection{Passage-Pointwise-Simple}

This instruction just provides simple task definition to rate the relevancy score.

\begin{tcolorbox}
Given a passage and a query, rate the relevancy level between the passage and the query from 0 to 5, where a higher score indicates larger relevancy. \\

Passage: \code{\{\{passage\}\}}

Query: \code{\{\{query\}\}}

Please rate the relevancy level between the passage and the query from 0 to 5.

Answer:
\end{tcolorbox}

\subsection{Passage-Pointwise-Complex}

This instruction provides complete task definition to rate the relevancy score.

\begin{tcolorbox}
This is the automatic relevancy evaluator of a retriever: \\
- Consider an input query and a corresponding passage \\
- Evaluate the passage according to one important quality: \\
1. Relevancy (0-5): a desired passage quality that requires the passage to include the answer of the query. \\
- All ratings are between 0-5 where 0 is very poor and 5 is very good.  \\
- The evaluation should be critical and careful, and should closely match the ratings of experts. This evaluation is very important.  \\
- Consider these aspects when evaluating: \\
    \indent 1. Query Understanding - Read the query carefully and understand the request of the query. \\
    \indent 2. Answer Finding - Read the passage and try finding the answer of the query from the passage. \\
    \indent 3. Assign a score for Relevancy on a scale of 0 to 5, where 0 is the lowest (hardest to find the answer of the query) and 5 is the highest (easiest to find the answer of the query) based on the Evaluation Criteria. \\

Given the passage and query, and prompt you to provide an evaluation. Respond with your integer 0-5 score first, then a rationale. \\

Passage: \code{\{\{passage\}\}} \\
Query: \code{\{\{query\}\}} \\
Relevancy Score:
\end{tcolorbox}

 \newpage
\subsection{Passage-Pointwise-Chat}

This instruction is used for black-box LLMs, such as GPT-3.5, GPT-4 and CLAUDE-3.5.

\begin{tcolorbox}
\textbf{system:}

You are the automatic relevancy evaluator of a retriever:  \\
- You consider an input query and a corresponding passage  \\
- You evaluate the passage according to one important quality:  \\
    1. Relevancy (0-5): a desired passage quality that requires the passage to include the answer of the query.  \\
- All ratings are between 0-5 where 0 is very poor and 5 is very good. \\
- Your evaluation should be critical and careful, and should closely match the ratings of experts. This evaluation is very important. \\
- Consider these aspects when evaluating: \\
    1. Query Understanding - Read the query carefully and understand the request of the query. \\
    2. Answer Finding - Read the passage and try finding the answer of the query from the passage. \\
    3. Assign a score for Relevancy on a scale of 0 to 5, where 0 is the lowest (hardest to find the answer of the query) and 5 is the highest (easiest to find the answer of the query) based on the Evaluation Criteria. \\

\textbf{user:} 

I will provide you with both the passage and query, and prompt you to provide an evaluation. Response with your integer 0-5 score first, then a rationale.

\textbf{assistant:}

Okay, please provide the passage and query.

\textbf{user:}

Passage: \code{\{\{passage\}\}} \\
Query: \code{\{\{query\}\}} \\
Relevancy Score:
\end{tcolorbox}

\subsection{Passage-Relwise}

This instruction is used for relevancy judgement.

\begin{tcolorbox}
Given a passage and a query, predict whether the passage includes an answer to the query by producing either 'Yes' or 'No'. \\

Passage: \code{\{\{passage\}\}} \\
Query: \code{\{\{query\}\}} \\
Does the passage answer the query? \\
Answer:
\end{tcolorbox}

\newpage
\subsection{Rank-Pointwise}

This instruction combines the complex task definition and rank representation.

\begin{tcolorbox}
This is the automatic relevancy evaluator of a retriever: \\
- Consider an input query and a rank of corresponding passages retrieved by the retriever \\
- Evaluate the rank of passages according to one important quality: \\
    1. Passage Relevancy: a desired passage quality that requires the passage to include the answer of the query. \\
    2. Rank Validity: a rank quality that increases the gain of passages ranked higher and reduce the loss of passages ranked lower \\
- All ratings are between 0-100 where 0 is very poor and 100 is very good. \\
- The evaluation should be critical and careful, and should closely match the ratings of experts. This evaluation is very important. \\
- Consider these aspects when evaluating: \\
    1. Query Understanding - Read the query carefully and understand the request of the query. \\
    2. Answer Finding - Read the passage and try finding the answer of the query from the passage. \\
    3. Assign a overall score for Passage Relevancy and Rank Validity on a scale of 0 to 100, where 0 is the lowest (hardest to find the answer of the query) and 100 is the highest (easiest to find the answer of the query) based on the Evaluation Criteria.

The following is the rank of 10 passages, each indicated by number identifier \textless\textgreater. The passages are listed in descending order using identifiers, and the most relevant passages should be listed first. \\

\textless1\textgreater  \code{\{\{passage\_1\}\}} \\
\textless2\textgreater  \code{\{\{passage\_2\}\}} \\
(more passages) ... \\

The search query is: \code{\{\{query\}\}} \\

Prompt you to provide an evaluation. Respond with your integer 0-100 score first, then a rationale. \\

The score of the rank is:
\end{tcolorbox}

\newpage
\subsection{Rank-Pairwise}

This instruction aims to compare the performance of two ranks.

\begin{tcolorbox}
This is RankEvaluator, an automatic evaluator that can evaluate the relevancy and quality of the assistants' ranked responses based on the query. \\

The following are two assistants' ranked responses, contained in the start and end identifier of respective assistant. In each ranked response, passages are listed in descending order using number identifiers \textless\textgreater, and the most relevant passages considered by respective assistant are listed first. I can evaluate the two ranked responsed based on their relevancy and quality to the query: \code{\{\{query\}\}}  \\

[The Start of Assistant 1's Ranked Response]

\textless1\textgreater  \code{\{\{passage\_1\}\}} \\
\textless2\textgreater  \code{\{\{passage\_2\}\}} \\
(more passages) ...

[The End of Assistant 1's Ranked Response] \\

[The Start of Assistant 2's Ranked Response]

\textless1\textgreater  \code{\{\{passage\_1\}\}} \\
\textless2\textgreater  \code{\{\{passage\_2\}\}} \\
(more passages) ... 

[The End of Assistant 2's Ranked Response] \\

The search query is: \code{\{\{query\}\}} \\

I will critically and carefully compare the quality of the above two assistants' ranked responses based on their relevancy to the search query. Select one assistant whose ranked response is more relevant and effective. \\

The more effective assistant is:
\end{tcolorbox}

\section{Complete results of prompting strategies}

The complete experimental results of different prompting strategies on all datasets and LLMs are shown in Table \ref{tab:complete_instructions}. It can be found that there is different effect of prompting strategies on datasets and LLMs. This is because their performances are affect by the style of dataset and LLMs' capacity. But, on the whole, the Passage-Pointwise-Complex has the overall best performance, which is selected as the default instruction on other experiments.

\begin{figure*}[t]
    \centering
    \includegraphics[width=0.9\linewidth]{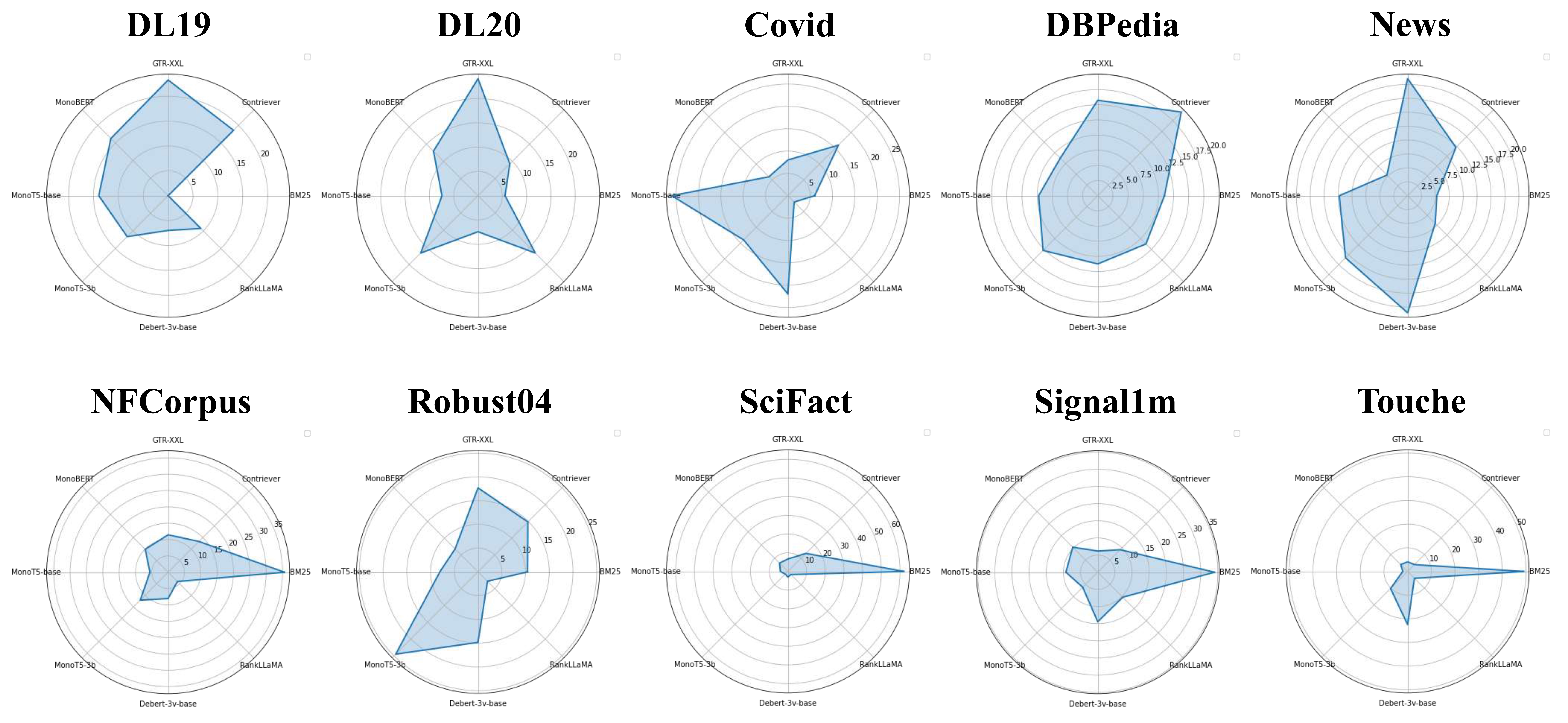}
    \caption{The frequency of base rerankers on each dataset of TREC and BEIR. The frequency definition is the same with Figure \ref{fig1}. These charts aim to reflect the preference of base rerankers on different datasets.}
    \label{fig4}
\end{figure*}

\begin{table*}[t]
    \centering
    \small
    \setlength\tabcolsep{4pt}
    \begin{tabular}{l cc | cccccccc | c }
         \toprule
         \textbf{Instructions} & DL19 &  DL20 & Covid &  DBPedia & News &  NFCorpus & Robust04 &  SciFact & Signal1m &  Touche & BEIR(Avg) \\
         \midrule
         \textbf{LLaMA3-8b} \\
         \midrule
         Passage-Pointwise-S & 72.65 & 70.59 & 82.24 & 46.68 & 46.09 & 36.14 & 57.02 & 72.06 & 29.98 & 30.66 & 50.11 \\ 
        Passage-Pointwise-C & 73.97 & 68.4 & 80.71 & 45.16 & 47.55 & 36.74 & 56.34 & 72.24 & 30.51 & 31.54 & 50.10 \\ 
        Passage-Relwise & 74.29 & 70.07 & 78.62 & 45.51 & 48.83 & 35.51 & 54.53 & 75.39 & 29.98 & 30.90 & 49.91 \\ 
        Rank-Pointwise & 72.83 & 68.93 & 80.71 & 44.43 & 48.47 & 37.07 & 56.02 & 76.54 & 31.28 & 32.77 & 50.91 \\ 
        Rank-Pairwise & 73.07 & 69.71 & 80.95 & 44.79 & 48.47 & 37.38 & 56.15 & 76.37 & 30.16 & 33.60 & 50.98 \\ 
        \midrule
         \textbf{LLaMA3-70b} \\
         \midrule
         Passage-Pointwise-C & 73.93 & 72.83 & 84.59 & 46.70 & 52.65 & 37.95 & 59.15 & 77.70 & 29.83 & 32.32 & 52.61 \\
        Passage-Relwise & 73.48 & 71.29 & 81.58 & 46.17 & 49.54 & 37.65 & 59.28 & 75.23 & 29.60 & 32.61 & 51.46 \\
        Rank-Pointwise & 72.81 & 68.89 & 80.71 & 44.33 & 48.47 & 37.58 & 56.56 & 76.59 & 31.74 & 32.41 & 51.05 \\
        Rank-Pairwise & 73.57 & 69.29 & 79.81 & 46.80 & 48.43 & 36.90 & 55.49 & 76.15 & 29.62 & 32.73 & 50.74 \\
        \midrule
         \textbf{Vicuna-v1.5-7b} \\
         \midrule
         Passage-Pointwise-S & 72.13 & 69.08 & 81.52 & 45.39 & 49.45 & 37.05 & 57.29 & 73.37 & 31.99 & 32.10 & 51.02 \\
        Passage-Pointwise-C & 73.73 & 70.46 & 80.69 & 45.95 & 49.23 & 36.33 & 56.52 & 72.55 & 29.94 & 29.84 & 50.13 \\
        Passage-Relwise & 74.27 & 74.07 & 78.06 & 45.91 & 50.43 & 36.74 & 58.75 & 72.94 & 28.67 & 30.33 & 50.23 \\
        Rank-Pointwise & 73.64 & 68.89 & 70.74 & 44.22 & 45.38 & 34.58 & 51.82 & 72.72 & 31.56 & 32.16 & 47.90 \\
        Rank-Pairwise & 71.68 & 68.23 & 76.18 & 44.91 & 45.00 & 35.48 & 49.58 & 70.65 & 32.32 & 31.51 & 48.20 \\
        \midrule
         \textbf{Vicuna-v1.5-13b} \\
         \midrule
        Passage-Pointwise-S & 73.84 & 74.02 & 76.68 & 47.63 & 50.92 & 37.92 & 60.48 & 75.68 & 29.64 & 31.05 & 51.25 \\
        Passage-Pointwise-C & 74.88 & 72.81 & 72.79 & 46.37 & 52.22 & 36.62 & 59.10 & 74.62 & 29.65 & 31.83 & 50.40 \\
        Passage-Relwise & 74.26 & 72.43 & 81.70 & 45.76 & 48.24 & 37.23 & 59.77 & 76.31 & 30.77 & 33.32 & 51.64 \\
        Rank-Pointwise & 73.79 & 69.00 & 71.40 & 43.68 & 48.75 & 34.93 & 52.56 & 72.73 & 31.05 & 28.14 & 47.91 \\
        Rank-Pairwise & 67.94 & 65.11 & 81.06 & 44.99 & 48.78 & 36.80 & 55.77 & 75.65 & 30.97 & 33.72 & 50.97 \\
        \midrule
         \textbf{ChatGLM-6b} \\
         \midrule
        Passage-Pointwise-S & 70.93 & 66.03 & 71.78 & 42.37 & 45.25 & 33.92 & 49.47 & 68.39 & 28.44 & 31.36 & 46.37 \\
        Passage-Pointwise-C & 66.7 & 64.75 & 65.38 & 40.31 & 42.89 & 33.51 & 43.92 & 67.87 & 27.69 & 27.19 & 43.60 \\
        Passage-Relwise & 73.22 & 69.93 & 82.48 & 44.63 & 48.09 & 36.84 & 55.96 & 73.43 & 31.96 & 31.14 & 50.57 \\
        Rank-Pointwise & 70.19 & 68.47 & 70.80 & 43.08 & 48.47 & 37.05 & 56.37 & 76.59 & 28.45 & 27.34 & 48.52 \\
        Rank-Pairwise & 67.88 & 64.58 & 79.72 & 41.62 & 48.47 & 37.07 & 55.98 & 76.59 & 32.25 & 32.68 & 50.55 \\
         \midrule
         \textbf{ChatGLM2-6b} \\
         \midrule
         Passage-Pointwise-S & 67.91 & 66.17 & 73.09 & 44.23 & 46.27 & 36.03 & 49.66 & 75.69 & 32.71 & 32.14 & 48.73 \\
        Passage-Pointwise-C & 64.65 & 64.17 & 71.07 & 42.05 & 44.05 & 34.12 & 48.21 & 70.21 & 30.54 & 33.00 & 46.66 \\
        Passage-Relwise & 74.38 & 70.03 & 80.14 & 45.51 & 46.77 & 36.72 & 57.10 & 75.67 & 31.70 & 31.56 & 50.65 \\
        Rank-Pointwise & 69.41 & 67.15 & 73.25 & 42.51 & 44.76 & 33.79 & 49.64 & 69.71 & 32.46 & 32.97 & 47.39 \\
        Rank-Pairwise & 70.74 & 68.78 & 67.96 & 41.09 & 42.78 & 33.01 & 47.77 & 67.37 & 29.32 & 32.58 & 45.24 \\
        \bottomrule

\end{tabular}
\caption{Results (nDCG@10) of all the intructions and LLMs on the TREC and BEIR. Passage-Pointwise-S means the Passage-Pointwise-Simple and Passage-Pointwise-C means the Passage-Pointwise-Complex.}
\label{tab:complete_instructions}
\end{table*}

\section{Complete frequency of selected retrieval models}

The effectiveness and potential analysis experiments count the frequency of all the selected retrieval models on each of datasets in the TREC and BEIR. The complete experimental results are shown in the Figure \ref{fig4}. It can be observed that there are clear differences of the models' performance among datasets, demonstrating no model can currently excel in all the queries and corpora. On the other hand, the models' behaviors is consistent under some scenarios. For example, BM25 exhibits the strong preference on the SciFact, Signal1m and Touche, which all are professional and long-context. This illustrates it have strong capacity upon those scenarios.

\end{document}